\documentclass{article}

\usepackage{arxiv}

\usepackage[utf8]{inputenc} 
\usepackage[T1]{fontenc}    
\usepackage{booktabs}       
\usepackage{amsfonts}       
\usepackage{nicefrac}       
\usepackage{microtype}      
\usepackage{lipsum}		
\usepackage{graphicx}
\usepackage{doi}

\title{Wide Field-of-View, Large-Area Long-wave Infrared Silicon Metalenses}

\author{Hung-I Lin\thanks{These authors contribute equally to this work}\\
Massachusetts Institute of Technology\\
Cambridge, MA \\
\And
Jeffrey Geldmeier\footnotemark[1] \\
Lockheed Martin Corporation\\
Orlando, FL \\
\And
Erwan Baleine\footnotemark[1] \\
Lockheed Martin Corporation\\
Orlando, FL \\
\And
Fan Yang\footnotemark[1] \\
Massachusetts Institute of Technology\\
Cambridge, MA \\
\And
Sensong An \\
Massachusetts Institute of Technology\\
Cambridge, MA \\
\And
Ying Pan \\
Massachusetts Institute of Technology\\
Cambridge, MA \\
\And
Clara Rivero-Baleine\thanks{Correspondence} \\
Lockheed Martin Corporation\\
Orlando, FL \\
\texttt{rivcar21@hotmail.com} \\
\And
Tian Gu\footnotemark[2] \\
Massachusetts Institute of Technology\\
Cambridge, MA \\
\texttt{gutian@mit.edu} \\
\And
Juejun Hu \\
Massachusetts Institute of Technology\\
Cambridge, MA \\
}

\begin{document}
\maketitle

\begin{abstract}
Long-wave infrared (LWIR, 8-12 $\mu m$ wavelengths) is a spectral band of vital importance to thermal imaging. Conventional LWIR optics made from single-crystalline Ge and chalcogenide glasses are bulky and fragile. The challenge is exacerbated for wide field-of-view (FOV) optics, which traditionally mandates multiple cascaded elements that severely add to complexity and cost. Here we designed and experimentally realized a LWIR metalens platform based on bulk Si wafers featuring 140$^\circ$ FOV. The metalenses, which have diameters exceeding 4 cm, were fabricated using a scalable wafer-level process involving photolithography and deep reactive ion etching. Using a metalens-integrated focal plane array, we further demonstrated wide-angle thermal imaging.
\end{abstract}

\keywords{Wide field-of-view optics, metasurface, metalens, LWIR, fisheye lens}

\section{Introduction}
LWIR, which coincides with the peak blackbody emission wavelengths of near-room-temperature objects, is strategically important to wide-ranging imaging applications spanning nigh vision, remote sensing, robotics, industrial process monitoring, building inspection, automotive sensing, gas detection, and beyond. Since most classical optical materials such as oxide glasses and polymers become opaque at LWIR due to phonon absorption, traditional LWIR optics resort to specialty materials such as single-crystalline Ge and chalcogenide glasses. These materials either incur a high cost to manufacture (Ge), or are mechanically fragile (chalcogenide glasses). Moreover, classical refractive optics made from these materials (Ge in particular) are temperature-sensitive due to thermo-optic focal drift. The challenges are further exacerbated when it comes to applications demanding a wide field-of-view (WFOV), since classical WFOV infrared optics entail a compound lens architecture comprising multiple (in general 4 or more) stacked optical elements to suppress coma aberration\cite{aburmad2014panoramic}. As a result, even LWIR lenses with a moderate FOV of around 60$^\circ$ each cost well above \$1,000 off-the-shelf.

Optical metasurfaces provide an alternative to classical refractive optics through modulation of the amplitude, phase, and polarization state of the wavefront using sub-wavelength nanostructures customarily termed meta-atoms \cite{yu2011light,ni2012broadband,capasso2018future,kamali2018review,aieta2012aberration,yang2023metasurface,west2014all,khorasaninejad2017metalenses,lalanne2017metalenses,tseng2018metalenses, gu2023reconfigurable}. While a large collection of metalenses have been implemented at visible and near-infrared wavelengths, relatively few demonstrations targeted the LWIR regime. Pioneering work by several groups have realized silicon-based LWIR metalenses \cite{fan2018high, huang2021long, li2022largest,hou2022lightweight}. Using Si as the metasurface material is advantageous in that it is amenable to large-area wafer-level manufacturing processes, and that deep reactive ion etching (DRIE) can produce high aspect ratio Si meta-atom structures ideal for large optical phase coverage and potentially dispersion engineering \cite{shan2022design}. Si wafers prepared using the common Czochralski method, however, are known to exhibit a strong optical absorption band centering at 9 $\mu m$ wavelength due to the presence of oxygen impurity \cite{conwell1952properties}. To mitigate the issue, Ge coupled with a ZnS antireflection layer has been adopted for metalens fabrication to suppress parasitic absorption across the LWIR band \cite{nalbant2022transmission}. The challenge of coma aberration suppression and expanding the FOV has nonetheless not been tackled by these pioneering investigations except in a recent report\cite{wirth2023large}. More recently, metalens arrays comprising five lens, each covering a sub-section of the FOV, have been implemented to demonstrate LWIR imaging spanning a horizontal FOV exceeding 60$^\circ$ upon image stitching in post-processing \cite{zhao2023wide}. The approach is however hardly scalable to WFOV applications, as a large FOV (e.g. 100$^\circ$) in both horizontal and vertical directions would require tens of individual metalenses, severely curtailing the optical throughput while escalating system complexity.

In this paper, we report the design and experimental demonstration of a WFOV metalens covering 140$^\circ$ circular FOV. The metalens assumed a simple architecture consisting of an optical aperture stop and a single-layer metasurface patterned in a float-zone Si wafer. The choice of float-zone Si contributes to suppression of the oxygen impurity absorption band while still enabling full leverage of industry-standard Si fabrication processes. Compared to other WFOV designs such as quadratic phase\cite{pu2017nanoapertures,martins2020metalenses,chen2020chip,zhang2020numerical,lassalle2021imaging,zhou2022metasurface,zhang2022design} and doublet metalenses\cite{arbabi2016miniature,groever2017meta,huang2021achromatic,martins2022fundamental,tang2020achromatic,kim2020doublet}, the present architecture is simple and yet does not compromise the imaging quality or optical efficiency \cite{yang2023wide}.

The rest of this paper is organized as follows. We will start with formulating the overarching analytical design approach. Two metalens designs were derived using the method, with an air gap and a ZnSe spacer, respectively. The former features a simpler construction whereas the latter has the advantage of enhanced FOV and imaging quality as predicted by our analytical theory and validated via numerical simulations. We then proceed to describe the fabrication protocols as well as experimental characterization of both metalenses.

\section{Analytical WFOV metalens design}
\begin{figure}[h!]
\centering\includegraphics[width=\linewidth]{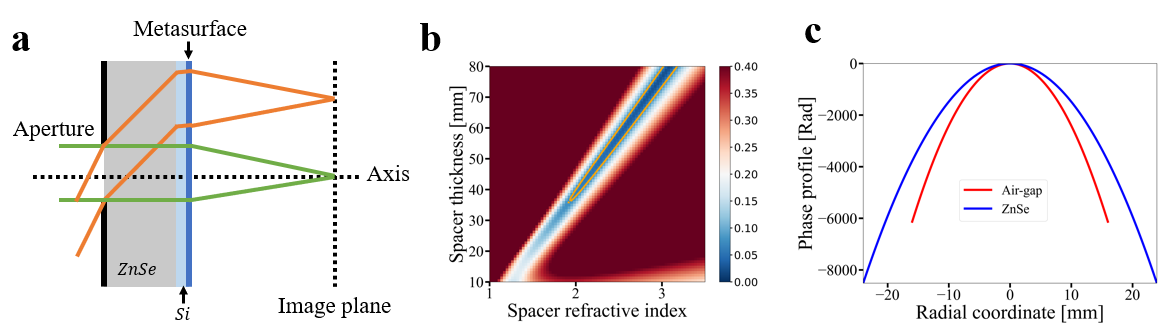}
\caption{Analytically guided WFOV metalens design optimization. (a) Schematic illustration of the WFOV metalens configuration. (b) Averaged RMS wavefront error across the FOV for a range of spacer refractive indices and thicknesses. The simulations assume 140$^\circ$ FOV and 20 $mm$ focal length. To highlight the design space relevant to high-performance imaging, RMS wavefront errors greater or equal to 0.4 $\lambda$ are represented in the plot using a uniform crimson color. The orange contour encircles the design parameter space where the RMS wavefront error is smaller than 0.0745 $\lambda$, which corresponds to the diffraction-limited performance criterion. (c) Phase profiles of the two metalens designs experimentally implemented in this study.}
\label{fig:1}
\end{figure}

The WFOV metalens architecture consisting of an aperture and an all-silicon metasurface is schematically illustrated in Fig. \ref{fig:1}a \cite{shalaginov2020single,yang2021design,shalaginov2022metasurface}. The monochromatic phase profiles of the metasurfaces are defined following \cite{yang2023understanding}:

\begin{equation}
\phi(s)=(\frac{2\pi}{\lambda})\int_0^s -(sin\alpha(s)+\frac{s-d}{\sqrt{f^2+(s-d)^2}})ds
\label{eq:1}
\end{equation}

The corresponding RMS wavefront error is given by:

\begin{equation}
\sigma\approx\frac{3nL^2D^3|s-d|}{160\left(f^2+\left(s-d\right)^2\right)(L^2+s^2)^{\frac{3}{2}}}
\label{eq:2}    
\end{equation}

We note that the numerator in the expression above contains the factor $|s-d|$, which corresponds to the transverse offset between the incident position of the chief ray on the metasurface and the corresponding focal spot position (i.e., image height). In an image-space telecentric configuration, the term vanishes, yielding optimal image quality. Optimizing the metalens design therefore involves engineering its image height vs. incident angle relation to mimic the telecentric configuration. This can be accomplished by changing the refractive index $n_{sub}$ of the spacer material, leveraging refraction at the air-spacer front surface as a practical means to modify the image height. Following this rationale, we examine the dependence of the RMS wavefront error for various spacer material refractive indices and thicknesses (Fig. \ref{fig:1}b).

Guided by the theoretical insight, we have chosen ZnSe ($n_{sub} = 2.40$ at 10.6 $\mu m$) as the spacer material. An air-gap design ($n_{sub} = 1$) was also implemented for comparison. The detailed design parameters are tabulated in Table \ref{tab:1} and Fig. \ref{fig:1}c plots the phase profiles of the designs.

\begin{table}[ht]
\caption{Metalens design parameters} 
\label{tab:1}
\begin{center} 
\begin{tabular}{|l|l|l|l|l|}
\hline
\rule[-1ex]{0pt}{3.5ex}  & Wavelength & Aperture size & Air-gap/ZnSe thickness & Si substrate thickness \\
\hline
\rule[-1ex]{0pt}{3.5ex} Air-gap & 10.6 $\mu m$ & 6 $mm$ & 7.5 $mm$ & 675 $\mu m$ \\
\hline
\rule[-1ex]{0pt}{3.5ex} ZnSe & 10.6 $\mu m$ & 10 $mm$ & 44 $mm$ & 675 $\mu m$ \\
\hline
\rule[-1ex]{0pt}{3.5ex} & Focal length & Metalens size & Image plane size & FOV \\
\hline
\rule[-1ex]{0pt}{3.5ex} Air-gap & 12 $mm$ & 32 $mm$ & 21 $mm$ & 90$^\circ$ \\
\hline
\rule[-1ex]{0pt}{3.5ex} ZnSe & 20 $mm$ & 48 $mm$ & 38 $mm$ & 140$^\circ$ \\
\hline
\end{tabular}
\end{center}
\end{table}

\begin{figure}[h!]
\centering\includegraphics[width=\linewidth]{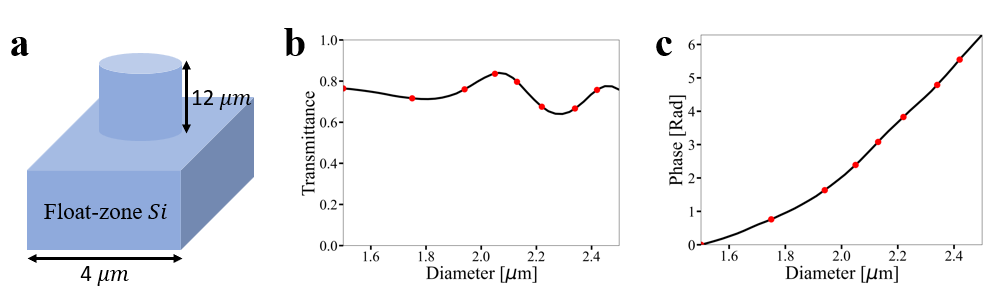}
\caption{All-Si meta-atom design. (a) Transmittance and (c) phase delay responses of the meta-atoms with different pillar diameters. Inset depicts the meta-atom structure and the red dots denote the eight meta-atom designs selected to construct the metasurfaces.}
\label{fig:2}
\end{figure}

Next we translate the phase functions into actual metasurface layouts. The all-Si meta-atom structure is depicted in Fig. \ref{fig:2}a inset, which is composed of 12 $\mu m$ tall pillars with a 4 $\mu m$ pitch etched into float-zone Si wafers. Full-wave electromagnetic simulations were performed using the Lumerical FDTD solver, and the transmittance and phase delay of the meta-atoms at 10.6 $\mu m$ wavelength as functions of the pillar diameter are shown in Figs. \ref{fig:2}a-b. Eight meta-atoms with approximately $\frac{\pi}{4}$ step size in phase were chosen to construct the metasurfaces. To optimize transmittance while suppressing phase error, we invoked a figure-of-merit function as the criterion to choose the meta-atom diameters \cite{shalaginov2021reconfigurable,yang2022reconfigurable}. The eight selected meta-atom designs are summarized in Table \ref{tab:2}.

\begin{table}[ht]
\caption{Meta-atom diameter, transmittance, and phase delay} 
\label{tab:2}
\begin{center} 
\begin{tabular}{|l|l|l|l|l|l|l|l|l|}
\hline
\rule[-1ex]{0pt}{3.5ex} Meta-atom index & 1 & 2 & 3 & 4 & 5 & 6 & 7 & 8 \\
\hline
\rule[-1ex]{0pt}{3.5ex} Phase [$^\circ$] & 0 & 44 & 94 & 137 & 176 & 219 & 274 & 318 \\
\hline
\rule[-1ex]{0pt}{3.5ex} Transmittance & 0.76 & 0.72 & 0.76 & 0.84 & 0.80 & 0.68 & 0.67 & 0.76 \\
\hline
\rule[-1ex]{0pt}{3.5ex} Diameter [$\mu m$] & 1.50 & 1.75 & 1.94 & 2.05 & 2.13 & 2.22 & 2.34 & 2.42 \\
\hline
\end{tabular}
\end{center}
\end{table}

\begin{figure}[h!]
\centering\includegraphics[width=\linewidth]{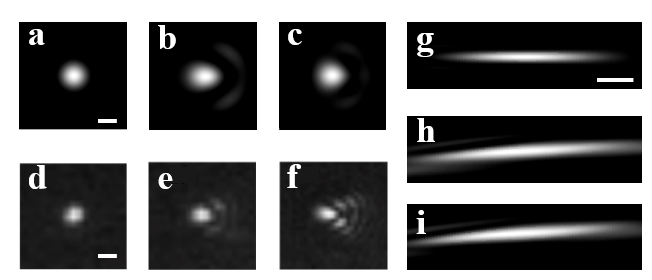}
\caption{Focusing characteristics of the air-gap WFOV metalens at 10.6 $\mu m$ wavelength. (a)-(c) Simulated PSFs of the metalens at its image plane for AOIs of (a) 0$^\circ$, (b) 20$^\circ$, and (c) 40$^\circ$. (Scale bar: 20 $\mu m$.) (d)-(f) Measured PSFs of the metalens at its image plane for AOIs of (d) 0$^\circ$, (e) 20$^\circ$, and (f) 40$^\circ$. (Scale bar: 20 $\mu m$.) (g)-(i) Simulated longitudinal intensity profiles of the metalens focal spot at AOIs of (g) 0$^\circ$, (h) 20$^\circ$, and (i) 40$^\circ$. (Scale bar: 100 $\mu m$.)}
\label{fig:3}
\end{figure}

\begin{figure}[h!]
\centering\includegraphics[width=\linewidth]{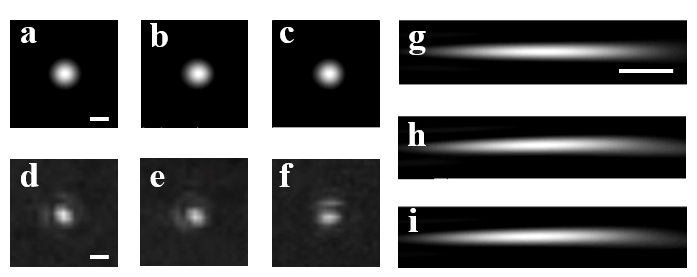}
\caption{Focusing characteristics of the ZnSe-spacer WFOV metalens at 10.6 $\mu m$ wavelength. (a)-(c) Simulated PSFs of the metalens at its image plane for AOIs of (a) 0$^\circ$, (b) 30$^\circ$, and (c) 70$^\circ$. (Scale bar: 20 $\mu m$.) (d)-(f) Measured PSFs of the metalens at its image plane for AOIs of (d) 0$^\circ$, (e) 30$^\circ$, and (f) 70$^\circ$. (Scale bar: 20 $\mu m$.) (g)-(i) Simulated longitudinal intensity profiles of the metalens focal spot at AOIs of (g) 0$^\circ$, (h) 30$^\circ$, and (i) 70$^\circ$. (Scale bar: 100 $\mu m$.)}
\label{fig:4}
\end{figure}

\begin{figure}[h!]
\centering\includegraphics[width=\linewidth]{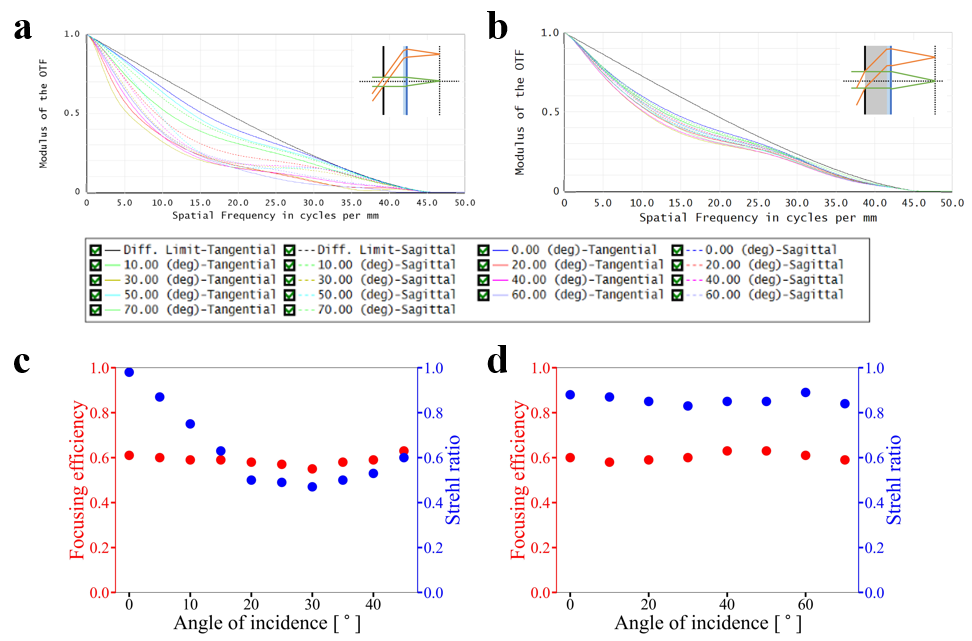}
\caption{Metalens performance at 10.6 $\mu m$ wavelength. (a)-(b) Simulated MTFs of the (a) air-gap and (b) ZnSe-spacer metalenses at different AOIs. (c)-(d) Focusing efficiency and Strehl ratio of the (c) air-gap and (d) ZnSe-spacer metalenses as functions of AOI.}
\label{fig:5}
\end{figure}

Based on the meta-atom characteristics, performances of the WFOV metalenses can be numerically evaluated using the Kirchhoff diffraction integral. The transverse and longitudinal focal spot intensity profiles of the two WFOV metalens designs at several different angles of incidence (AOIs) are presented in Fig. \ref{fig:3} and Fig. \ref{fig:4}, respectively. The modulation transfer functions (MTFs) at different spatial frequencies were obtained through Fourier transform of the simulated point-spread-functions (PSFs) and are shown in Figs. \ref{fig:5}a-b.

We now characterize the focusing efficiencies and Strehl ratios of the WFOV metalenses at 10.6 $\mu m$ wavelength using the numerical results in Figs. \ref{fig:3} and \ref{fig:4}. Here the focusing efficiency is defined as the fraction of power encircled within a diameter equaling five times the focal spot full-width-at-half-maximum (FWHM), normalized by the total incident power \cite{yang2021design}. Figures \ref{fig:5}c-d plot the two parameters as functions of AOI. The air-gap metalens has a focusing efficiency of 53$\%$ and a Strehl ratio of 0.63, both averaged over AOIs across the 90$^\circ$ FOV, whereas the ZnSe-spacer metalens claims a focusing efficiency of 50$\%$ and a Strehl ratio of 0.86, similarly averaged over AOIs throughout the entire 140$^\circ$ FOV. The enhanced focusing performance of the ZnSe-spacer lens over the air-gap design, evidenced by its diffraction-limited performance (Strehl ratio $>$ 0.8) over an extended FOV of 140$^\circ$, validates our theoretical prediction.

\section{Metalens fabrication}
\begin{figure}[h!]
\centering\includegraphics[width=\linewidth]{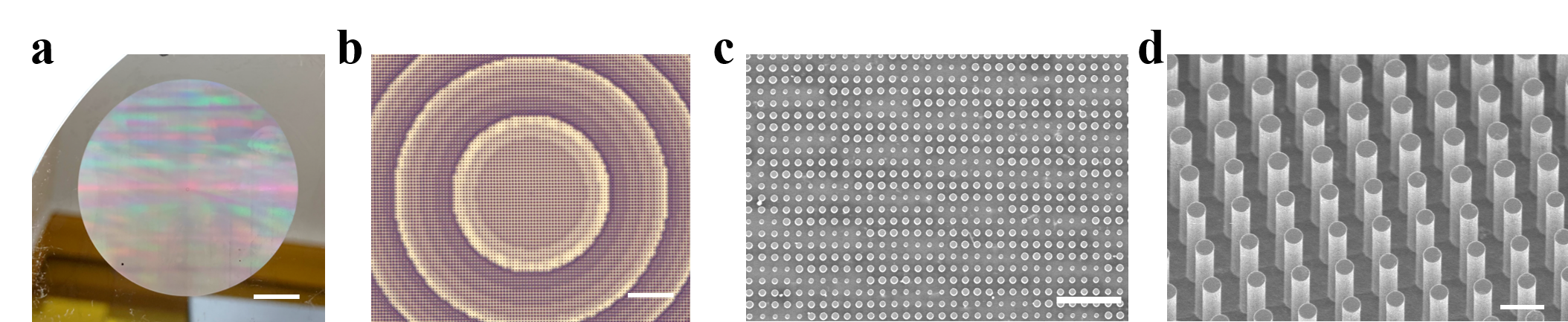}
\caption{Images of fabricated metasurfaces. (a) A photo of a metasurface sample. (Scale bar: 10 mm.) (b) An optical micrograph of the metasurface. (Scale bar: 60 $\mu m$.) (c) Top-view SEM image of the meta-atoms. (Scale bar: 20 $\mu m$.) (d) Tilted-view SEM image showing scalloping-free sidewall profiles of the meta-atoms. (Scale bar: 4 $\mu m$.)}
\label{fig:6}
\end{figure}

2 $\mu m$ thick SiN films were deposited by plasma-enhanced chemical vapor deposition (STS PECVD) on 675 $\mu m$ thick float zone Si wafers as a hard mask for DRIE. To define the metasurface patterns, a negative-tone photoresist (AZ nLOF 2035) was spin-coated onto the substrates at 3000 revolutions per minute (rpm). The resist was soft-baked at 115 $^\circ$C for 1 minute, exposed on an MLA150 Maskless Aligner, and then post-exposure baked at 115 $^\circ$C for 1 minute. The photoresist was developed by immersing the sample in Microposit MF-319 developer for 1 minute, followed by rinsing in deionized water. To etch the SiN hard mask, dry etching was performed using dual gas inlets with a mixture of SF\textsubscript{6} and C\textsubscript{4}F\textsubscript{8} (STS ICP RIE). The Bosch process was subsequently used to etch the Si meta-atoms (SPTS Rapier DRIE) before hard mask removal via buffered HF (BHF) wet etching to complete the fabrication process. To mitigate scalloping which is commonly associated with the Bosch process, we optimize 1) the gas flow ratio, which balances the etching and passivation effects; and 2) the etching loop time, which dictates the spatial period and severity of scalloping (Supplementary Information). Figure \ref{fig:6} shows images of the metasurfaces showing a low-roughness sidewall profile with minimal scalloping fabricated using the optimized parameters.

\section{Metalens characterization and thermal imaging demonstration}

\begin{figure}[h!]
\centering\includegraphics[width=\linewidth]{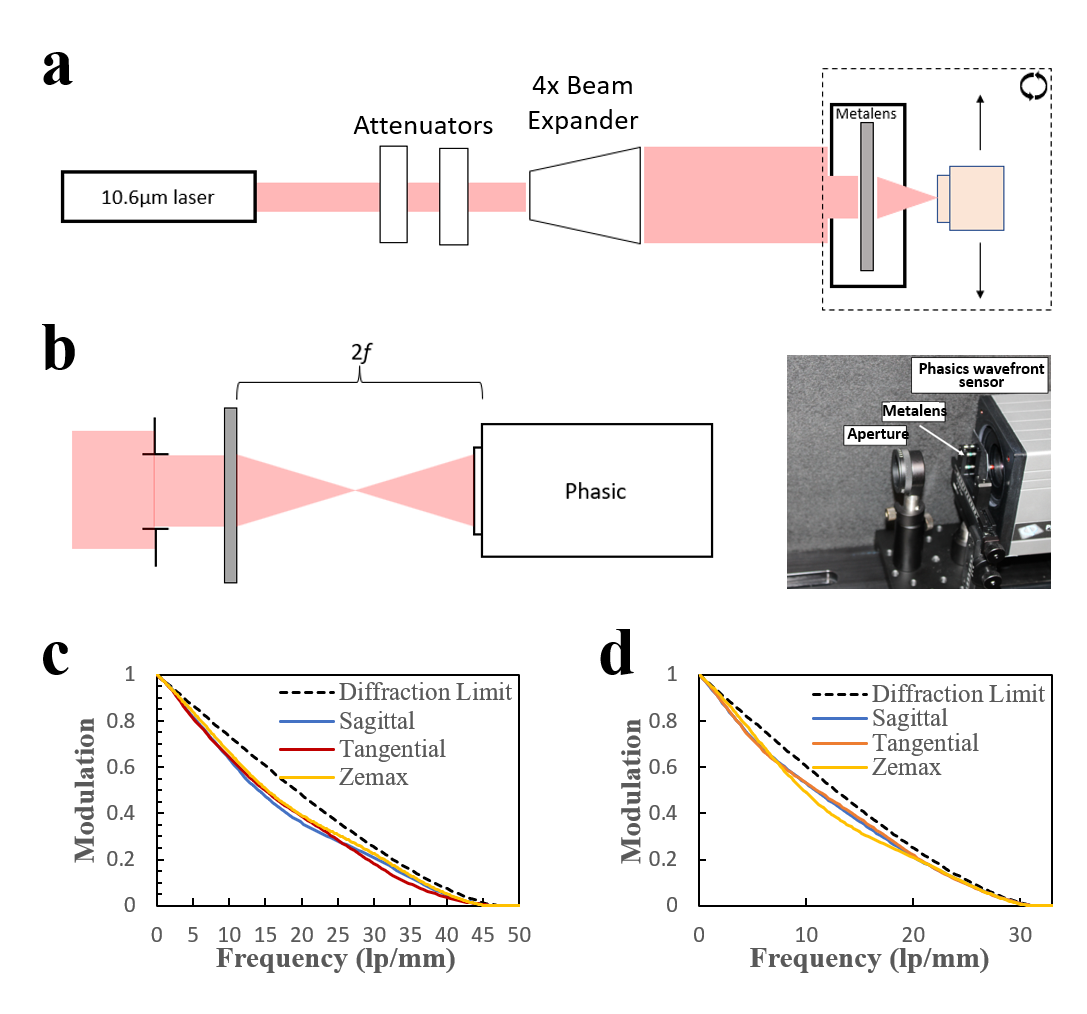}
\caption{Experimental PSF and MTF measurements at 10.6 $\mu m$ wavelength. (a) Experimental setup for characterizing the metalens PSFs at different AOIs. (b) Experimental setup for wavefront measurement. (c)-(d) Normal-incidence MTFs of the (c) air-gap and (d) ZnSe metalens based on the wavefront measurement.}
\label{fig:7}
\end{figure}

The PSFs of the metalenses were characterized experimentally using a setup schematically depicted in Fig. \ref{fig:7}a. A collimated 10.6 $\mu m$ CO\textsubscript{2} laser (L4GST, Access) was first attenuated and then expanded before being directed at the metalens and subsequently focused onto a microbolometer (Boson, FLIR). For field angles away from $0^\circ$, the metalens and the detector were tilted together, and the microbolometer also required lateral movement to span the much larger image plane. Several examples of the metalens PSFs at different AOIs are presented in Figs. \ref{fig:4}d-f and Figs. \ref{fig:5}d-f, respectively, showing excellent agreement between the measurement results and simulations. The monochromatic MTF performance of the metalenses at 10.6 $\mu m$ was measured using the collimated CO\textsubscript{2} laser and an interferometric wavefront sensor (SID4 DWIR, Phasics) schematically depicted in Fig. \ref{fig:7}b. An aperture was placed in front of the metalenses in order to limit the collimated light bundle to the entrance pupil size. The Phasics camera was set to the focal mode, allowing measurements to be collected after direct diffraction from the lens. For the ZnSe metalens, an additional pinhole was placed near the focal plane to remove the zero order component. The corresponding MTFs are shown in Fig. \ref{fig:7}c-d, indicating diffraction-limited performance in agreement with simulation predictions.

\begin{figure}[h!]
\centering\includegraphics[width=\linewidth]{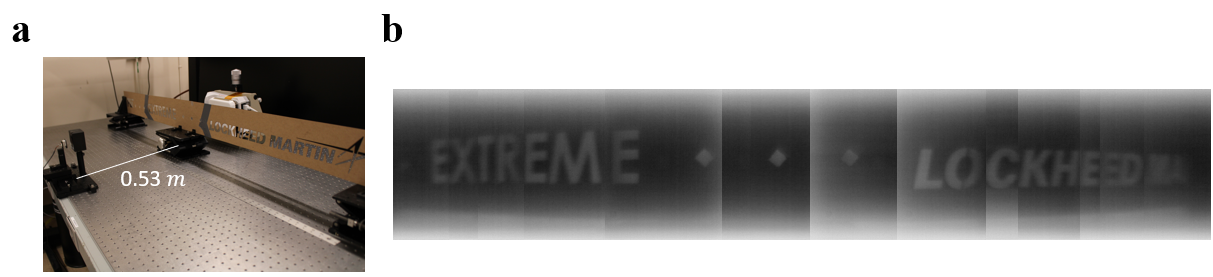}
\caption{Wide-angle thermal imaging. (a) Experimental setup of the thermal imaging test. (b) Images taken by the thermal infrared camera integrated with the ZnSe-spacer metalens. Since the image size is larger than the bolometer sensor area, the bolometer was laterally translated with respect to the metalens and the resulting image sections were stitched.}
\label{fig:8}
\end{figure}

Finally, we demonstrated wide-angle thermal imaging using the metalens. The ZnSe metalens was mounted onto a microbolometer (Boson, FLIR) to form a thermal infrared camera. The experimental setup is shown in Fig. \ref{fig:8}a. A hot plate was placed 0.53 meter away in front of the WFOV metalens to act as a LWIR illumination source. A card board 1.26 meters in length and perforated with 'EXTREME LOCKHEED MARTIN' patterns was placed in between the hot plate and the metalens, which blocked the blackbody radiation from the hot plate in all areas except within the inverse 'EXTREME LOCKHEED MARTIN' pattern. A 10.5 $\mu m$ filter with 0.2 $\mu m$ bandwidth was placed in front of the sensor. The captured image is presented in Fig. \ref{fig:8}b covering 100$^\circ$ FOV. Since the image size is larger than the image sensor area, the bolometer had to be laterally translated with respect to the metalens and the resulting image sections were stitched to form Fig. \ref{fig:8}b. We further compared the imaging results with those using an unfiltered blackbody source to quantify the impact of chromatic aberration, and the details were elaborated in the Supplementary Information.

\section{Conclusion}

In this paper, we reported the design and experimental demonstration of metalenses operating at the LWIR band with an ultra-wide FOV of 140$^\circ$. An analytical model was used to rationally guide the design of the metasurfaces as well as the lens spacer material choice. Following the designs, centimeter-scale metasurfaces were fabricated on float-zone silicon wafers using large-area photolithographic patterning and optimized DRIE protocols. Experimental characterization of the metalenses validated our theoretical design and demonstrated thermal imaging using a metalens-integrated infrared camera. Benefiting from its simple architecture, scalable fabrication process, and exceptional wide-FOV imaging capability, the WFOV metalens technology potentially offers an appealing alternative to existing LWIR compound lens optics for applications covering night vision, spectroscopic sensing, environmental monitoring, and many others.

\section{Acknowledgements}
The authors acknowledge nanofabrication and characterization facility support provided by MIT.nano and the Center for Nanoscale Systems at Harvard University.

\section{Funding}
This work was supported by Lockheed Martin Corporation Internal Research and Development and Defense Advanced Research Projects Agency Defense Sciences Office Program: EXTREME Optics and Imaging (EXTREME) under agreement number HR00111720029. The views, opinions and/or findings expressed are those of the authors and should not be interpreted as representing the official views or policies of the Department of Defense or the US Government.

\bibliographystyle{unsrt}
\bibliography{main}      

\begin{thebibliography}{10}

\bibitem{aburmad2014panoramic}
Shimon Aburmad.
\newblock Panoramic thermal imaging: challenges and tradeoffs.
\newblock In {\em Infrared Technology and Applications XL}, volume 9070, pages
  115--123. SPIE, 2014.

\bibitem{yu2011light}
Nanfang Yu, Patrice Genevet, Mikhail~A Kats, Francesco Aieta, Jean-Philippe
  Tetienne, Federico Capasso, and Zeno Gaburro.
\newblock Light propagation with phase discontinuities: generalized laws of
  reflection and refraction.
\newblock {\em science}, 334(6054):333--337, 2011.

\bibitem{ni2012broadband}
Xingjie Ni, Naresh~K Emani, Alexander~V Kildishev, Alexandra Boltasseva, and
  Vladimir~M Shalaev.
\newblock Broadband light bending with plasmonic nanoantennas.
\newblock {\em Science}, 335(6067):427--427, 2012.

\bibitem{capasso2018future}
Federico Capasso.
\newblock The future and promise of flat optics: a personal perspective.
\newblock {\em Nanophotonics}, 7(6):953--957, 2018.

\bibitem{kamali2018review}
Seyedeh~Mahsa Kamali, Ehsan Arbabi, Amir Arbabi, and Andrei Faraon.
\newblock A review of dielectric optical metasurfaces for wavefront control.
\newblock {\em Nanophotonics}, 7(6):1041--1068, 2018.

\bibitem{aieta2012aberration}
Francesco Aieta, Patrice Genevet, Mikhail~A Kats, Nanfang Yu, Romain Blanchard,
  Zeno Gaburro, and Federico Capasso.
\newblock Aberration-free ultrathin flat lenses and axicons at telecom
  wavelengths based on plasmonic metasurfaces.
\newblock {\em Nano letters}, 12(9):4932--4936, 2012.

\bibitem{yang2023metasurface}
Fan Yang, Hung-I Lin, Mikhail Shalaginov, Juejun Hu, and Tian Gu.
\newblock Metasurface optics enabled computational sensing.
\newblock In {\em AI and Optical Data Sciences IV}, volume 12438, pages 60--67.
  SPIE, 2023.

\bibitem{west2014all}
Paul~R West, James~L Stewart, Alexander~V Kildishev, Vladimir~M Shalaev,
  Vladimir~V Shkunov, Friedrich Strohkendl, Yuri~A Zakharenkov, Robert~K Dodds,
  and Robert Byren.
\newblock All-dielectric subwavelength metasurface focusing lens.
\newblock {\em Optics express}, 22(21):26212--26221, 2014.

\bibitem{khorasaninejad2017metalenses}
Mohammadreza Khorasaninejad and Federico Capasso.
\newblock Metalenses: Versatile multifunctional photonic components.
\newblock {\em Science}, 358(6367):eaam8100, 2017.

\bibitem{lalanne2017metalenses}
Philippe Lalanne and Pierre Chavel.
\newblock Metalenses at visible wavelengths: past, present, perspectives.
\newblock {\em Laser \& Photonics Reviews}, 11(3):1600295, 2017.

\bibitem{tseng2018metalenses}
Ming~Lun Tseng, Hui-Hsin Hsiao, Cheng~Hung Chu, Mu~Ku Chen, Greg Sun, Ai-Qun
  Liu, and Din~Ping Tsai.
\newblock Metalenses: advances and applications.
\newblock {\em Advanced Optical Materials}, 6(18):1800554, 2018.

\bibitem{gu2023reconfigurable}
Tian Gu, Hyun~Jung Kim, Clara Rivero-Baleine, and Juejun Hu.
\newblock Reconfigurable metasurfaces towards commercial success.
\newblock {\em Nature Photonics}, 17(1):48--58, 2023.

\bibitem{fan2018high}
Qingbin Fan, Mingze Liu, Cheng Yang, Le~Yu, Feng Yan, and Ting Xu.
\newblock A high numerical aperture, polarization-insensitive metalens for
  long-wavelength infrared imaging.
\newblock {\em Applied Physics Letters}, 113(20):201104, 2018.

\bibitem{huang2021long}
Luocheng Huang, Zachary Coppens, Kent Hallman, Zheyi Han, Karl~F B{\"o}hringer,
  Neset Akozbek, Ashok Raman, and Arka Majumdar.
\newblock Long wavelength infrared imaging under ambient thermal radiation via
  an all-silicon metalens.
\newblock {\em Optical Materials Express}, 11(9):2907--2914, 2021.

\bibitem{li2022largest}
Junwei Li, Yilin Wang, Shengjie Liu, Ting Xu, Kai Wei, Yudong Zhang, and Hao
  Cui.
\newblock Largest aperture metalens of high numerical aperture and polarization
  independence for long-wavelength infrared imaging.
\newblock {\em Optics Express}, 30(16):28882--28891, 2022.

\bibitem{hou2022lightweight}
Mingming Hou, Yan Chen, and Fei Yi.
\newblock Lightweight long-wave infrared camera via a single
  5-centimeter-aperture metalens.
\newblock In {\em CLEO: QELS\_Fundamental Science}, page FM4F.4. Optica
  Publishing Group, 2022.

\bibitem{shan2022design}
Dongzhi Shan, Nianxi Xu, Jinsong Gao, Naitao Song, Hai Liu, Yang Tang, Xiaoguo
  Feng, Yansong Wang, Yi~Zhao, Xin Chen, et~al.
\newblock Design of the all-silicon long-wavelength infrared achromatic
  metalens based on deep silicon etching.
\newblock {\em Optics Express}, 30(8):13616--13629, 2022.

\bibitem{conwell1952properties}
Esther~M Conwell.
\newblock Properties of silicon and germanium.
\newblock {\em Proceedings of the IRE}, 40(11):1327--1337, 1952.

\bibitem{nalbant2022transmission}
Halil~Can Nalbant, Fatih Balli, Tolga Yelbo{\u{g}}a, Arda Eren, and Ahmet
  S{\"o}zak.
\newblock Transmission optimized lwir metalens.
\newblock {\em Applied Optics}, 61(33):9946--9950, 2022.

\bibitem{wirth2023large}
Anna Wirth-Singh, Johannes~E Fr{\"o}ch, Zheyi Han, Luocheng Huang, Saswata
  Mukherjee, Zhihao Zhou, Zachary Coppens, Karl~F B{\"o}hringer, and Arka
  Majumdar.
\newblock Large field-of-view thermal imaging via all-silicon meta-optics.
\newblock {\em arXiv preprint arXiv:2304.14569}, 2023.

\bibitem{zhao2023wide}
Chenglong Zhao, Ziqi Liu, and Wei Huang.
\newblock Wide field-of-view metalens array for the long-wavelength infrared.
\newblock In {\em Conference on Infrared, Millimeter, Terahertz Waves and
  Applications (IMT2022)}, volume 12565, pages 746--749. SPIE, 2023.

\bibitem{pu2017nanoapertures}
Mingbo Pu, Xiong Li, Yinghui Guo, Xiaoliang Ma, and Xiangang Luo.
\newblock Nanoapertures with ordered rotations: symmetry transformation and
  wide-angle flat lensing.
\newblock {\em Optics Express}, 25(25):31471--31477, 2017.

\bibitem{martins2020metalenses}
Augusto Martins, Kezheng Li, Juntao Li, Haowen Liang, Donato Conteduca,
  Ben-Hur~V Borges, Thomas~F Krauss, and Emiliano~R Martins.
\newblock On metalenses with arbitrarily wide field of view.
\newblock {\em ACS Photonics}, 7(8):2073--2079, 2020.

\bibitem{chen2020chip}
Cong Chen, Panpan Chen, Jianxin Xi, Wanxia Huang, Kuanguo Li, Li~Liang, Fenghua
  Shi, and Jianping Shi.
\newblock On-chip monolithic wide-angle field-of-view metalens based on
  quadratic phase profile.
\newblock {\em AIP Advances}, 10(11):115213, 2020.

\bibitem{zhang2020numerical}
Wen-peng Zhang, Fei Liang, Ya-rong Su, Ke~Liu, Ming-jun Tang, Ling Li,
  Zheng-wei Xie, and Wu-ming Liu.
\newblock Numerical simulation research of wide-angle beam steering based on
  catenary shaped ultrathin metalens.
\newblock {\em Optics Communications}, 474:126085, 2020.

\bibitem{lassalle2021imaging}
Emmanuel Lassalle, Tobias~WW Mass, Damien Eschimese, Anton~V Baranikov, Egor
  Khaidarov, Shiqiang Li, Ramon Paniagua-Dominguez, and Arseniy~I Kuznetsov.
\newblock Imaging properties of large field-of-view quadratic metalenses and
  their applications to fingerprint detection.
\newblock {\em ACS Photonics}, 8(5):1457--1468, 2021.

\bibitem{zhou2022metasurface}
Guangzhu Zhou, Shi-Wei Qu, Baojie Chen, Yuansong Zeng, and Chi~Hou Chan.
\newblock Metasurface-based fourier lens fed by compact plasmonic optical
  antennas for wide-angle beam steering.
\newblock {\em Optics Express}, 30(12):21918--21930, 2022.

\bibitem{zhang2022design}
Ning Zhang, Qingzhi Li, Jun Chen, Feng Tang, Jingjun Wu, Xin Ye, and Liming
  Yang.
\newblock Design of an all-dielectric long-wave infrared wide-angle metalens.
\newblock {\em Chinese Physics B}, 31(7):074212, 2022.

\bibitem{arbabi2016miniature}
Amir Arbabi, Ehsan Arbabi, Seyedeh~Mahsa Kamali, Yu~Horie, Seunghoon Han, and
  Andrei Faraon.
\newblock Miniature optical planar camera based on a wide-angle metasurface
  doublet corrected for monochromatic aberrations.
\newblock {\em Nature Communications}, 7:13682, 2016.

\bibitem{groever2017meta}
Benedikt Groever, Wei~Ting Chen, and Federico Capasso.
\newblock Meta-lens doublet in the visible region.
\newblock {\em Nano letters}, 17(8):4902--4907, 2017.

\bibitem{huang2021achromatic}
Zhenyu Huang, Maosen Qin, Xiaowei Guo, Cheng Yang, and Shaorong Li.
\newblock Achromatic and wide-field metalens in the visible region.
\newblock {\em Optics Express}, 29(9):13542--13551, 2021.

\bibitem{martins2022fundamental}
Augusto Martins, Juntao Li, Ben-Hur~V Borges, Thomas~F Krauss, and Emiliano~R
  Martins.
\newblock Fundamental limits and design principles of doublet metalenses.
\newblock {\em Nanophotonics}, 11(6):1187--1194, 2022.

\bibitem{tang2020achromatic}
Dongliang Tang, Long Chen, Jia Liu, and Xiaohu Zhang.
\newblock Achromatic metasurface doublet with a wide incident angle for light
  focusing.
\newblock {\em Optics Express}, 28(8):12209--12218, 2020.

\bibitem{kim2020doublet}
Changhyun Kim, Sun-Je Kim, and Byoungho Lee.
\newblock Doublet metalens design for high numerical aperture and simultaneous
  correction of chromatic and monochromatic aberrations.
\newblock {\em Optics Express}, 28(12):18059--18076, 2020.

\bibitem{yang2023wide}
Fan Yang, Mikhail~Y Shalaginov, Hung-I Lin, Sensong An, Anu Agarwal, Hualiang
  Zhang, Clara Rivero-Baleine, Tian Gu, and Juejun Hu.
\newblock Wide field-of-view metalens: a tutorial.
\newblock {\em Advanced Photonics}, 5(3):033001--033001, 2023.

\bibitem{shalaginov2020single}
Mikhail~Y Shalaginov, Sensong An, Fan Yang, Peter Su, Dominika Lyzwa,
  Anuradha~M Agarwal, Hualiang Zhang, Juejun Hu, and Tian Gu.
\newblock Single-element diffraction-limited fisheye metalens.
\newblock {\em Nano Letters}, 20(10):7429--7437, 2020.

\bibitem{yang2021design}
Fan Yang, Sensong An, Mikhail~Y Shalaginov, Hualiang Zhang, Clara
  Rivero-Baleine, Juejun Hu, and Tian Gu.
\newblock Design of broadband and wide-field-of-view metalenses.
\newblock {\em Optics Letters}, 46(22):5735--5738, 2021.

\bibitem{shalaginov2022metasurface}
Mikhail~Y Shalaginov, Hung-I Lin, Fan Yang, Drew~M Weninger, Crystal Li,
  Anuradha~M Agarwal, Juejun Hu, and Tian Gu.
\newblock Metasurface-enabled wide-angle stereoscopic imaging.
\newblock In {\em Frontiers in Optics}, pages JTu7B--2. Optica Publishing
  Group, 2022.

\bibitem{yang2023understanding}
Fan Yang, Sensong An, Mikhail~Y Shalaginov, Hualiang Zhang, Juejun Hu, and Tian
  Gu.
\newblock Understanding wide field-of-view flat lenses: an analytical solution.
\newblock {\em Chinese Optics Letters}, 21(2):023601, 2023.

\bibitem{shalaginov2021reconfigurable}
Mikhail~Y Shalaginov, Sensong An, Yifei Zhang, Fan Yang, Peter Su, Vladimir
  Liberman, Jeffrey~B Chou, Christopher~M Roberts, Myungkoo Kang, Carlos Rios,
  et~al.
\newblock Reconfigurable all-dielectric metalens with diffraction-limited
  performance.
\newblock {\em Nature communications}, 12(1):1225, 2021.

\bibitem{yang2022reconfigurable}
Fan Yang, Hung-I Lin, Mikhail~Y Shalaginov, Katherine Stoll, Sensong An, Clara
  Rivero-Baleine, Myungkoo Kang, Anuradha Agarwal, Kathleen Richardson,
  Hualiang Zhang, et~al.
\newblock Reconfigurable parfocal zoom metalens.
\newblock {\em Advanced Optical Materials}, 10(17):2200721, 2022.

\end{thebibliography}
\end{document}